\documentclass[runningheads]{llncs}
\usepackage{graphicx}
\usepackage{algorithm}
\usepackage[noend]{algpseudocode}
\algnewcommand\And{\textbf{ and }}
\algnewcommand\Not{\textbf{not }}
\usepackage{xcolor}
\definecolor{listinggray}{gray}{0.9}
\definecolor{lbcolor}{rgb}{0.9,0.9,0.9}
\usepackage{listings}
\usepackage{amsfonts}

\usepackage{amsmath}

\begin{document}
\title{Towards the Democratization and Standardization of Dynamic Resources with MPI Spawning}
\titlerunning{Democratizing MPI Spawn-based Dynamic Resources}
%
\author{Sergio Iserte\inst{1}\orcidID{0000-0003-3654-7924} \and
Iker Martín-Álvarez\inst{2}\orcidID{0000-0002-3337-3298} \and
Krzystof Rojek\inst{3}\orcidID{0000-0002-2635-7345} \and
José I. Aliaga\inst{2}\orcidID{0000-0001-8469-764X} \and
Maribel Castillo\inst{2}\orcidID{0000-0002-2826-3086} \and \\
Antonio J. Peña\inst{1}\orcidID{0000-0002-3575-4617}}
\authorrunning{S. Iserte et al.}
%
\institute{Barcelona Supercomputing Center, Spain
\email{\{siserte, antonio.pena\}@bsc.es}\\
\and
Universitat Jaume I, Spain
\email{\{martini, aliaga, castillo\}@uji.es}\\
\and
Czestochowa University of Technology, Poland 
\email{krojek@icis.pcz.pl}}
\maketitle              
\begin{abstract}
This paper presents an efficient tool for managing dynamic resources in production high-performance computing (HPC) settings, focusing on flexibility, adaptability, and user-friendliness. We introduce a unified dynamic resource management application programming interface (API) that supports a wide range of HPC applications, allowing seamless integration without direct interaction with Dynamic Management of Resources (DMR). The DMR framework, evolved from the DMRlib structure, now supports various dynamic resource managers and includes the Proteo reconfiguration engine to enhance malleability strategies. This integration addresses previous limitations by allowing diverse reconfiguration methods without respawning all processes or lacking RMS support. 
The paper also showcases the solution's performance and coding productivity with the MPDATA (Multidimensional Positive Definite Advection Transport Algorithm) application. Key contributions include an enhanced modular DMR framework supporting different reconfiguration managers, upgraded DMRlib with the Proteo reconfiguration engine, offering extensive reconfiguration strategies, and a malleable version of the MPDATA solver.

\end{abstract}
%


\section{Introduction}
In recent years, a resurgence in dynamic resource management in high-performance computing (HPC) has emerged, driven by advancements in techniques, increased funding, and evolving needs.
Regarding techniques, resource manager systems (RMS) like Flux or OAR have emerged to challenge the globally dominant Slurm, asserting their readiness to handle workloads in the Exaflop era efficiently.
Moreover, new standards of MPI have presented more scalable tools for managing massive processes.
In terms of funding, the most recent (2019) EuroHPC-Joint Undertaking (JU) investment round supported various European projects,  including developing a dynamic resource manager tailored to each project's requirements. As institutions and users increasingly strive to enhance scientific output and maximize resource utilization, dynamic resource management has become a relevant technique to implement in production environments.

This paper aims to offer an efficient tool for managing dynamic resources in production HPC environments, demonstrating its flexibility, adaptability, and user-friendly interface. 
To achieve this goal, we have developed a unified dynamic resource management application programming interface (API) designed to cater to a vast variety of HPC applications, improving their flexible implementation. 
This API enables specific cases to seamlessly utilize the dynamic resource manager without the need for direct interaction with the DMR, the key focus of our research.

The DMR framework has evolved from the original DMRlib structure\cite{Iserte2020}, which initially provided syntax and technical support to facilitate adaptability.
As a result, DMR is now capable of accommodating various dynamic resource managers. Notably, DMRlib has transitioned into a standalone manager.
This transformation has enabled the DMRlib architecture to seamlessly substitute reconfiguration engines with minimal effort. Specifically, our work allows the use of the Proteo reconfiguration engine\cite{proteo_2024} within DMRlib to benefit from its diverse malleability strategies.
Through this combination, we have addressed the limitations of DMRlib and Proteo when used independently: the former's reliance on a single operating mode where all processes are re-spawned during each reconfiguration and the latter's lack of RMS support.
This paper advances state-of-the-art by presenting the new DMR modular architecture and demonstrating how it can adopt other powerful reconfiguration engines by establishing communication with the RMS and providing a common syntax for developers.

This integration demonstrates the versatility of DMR to be extended with results in a functional dynamic resource manager 

Finally, we evaluate the solution's performance and the coding productivity with a long-haul scientific application such as MPDATA (Multidimensional Positive Definite Advection Transport Algorithm).

In summary, the main contributions of this paper are:
\begin{enumerate}
    \item An enhanced stacked architecture for managing dynamic resources capable of supporting different reconfiguration managers.
    \item Proteo provides different spawning strategies and data redistribution methods that can now be used in production systems thanks to the integration in DMR.
    \item The proof that the revisited syntax simplifies the development of malleable applications with the example of MPDATA.
\end{enumerate}

The rest of the paper is structured as follows: 
Section~\ref{sec:background} thoroughly describes the technologies used in this research. 
Section~\ref{sec:related} discusses the related work in the area of MPI malleability and dynamic resource management.
Section~\ref{sec:dmr} explains how the original DMRlib has been split into DMR and the new DMRlib.
Section~\ref{sec:proteo} details the steps followed to use Proteo within DMRlib.
In Section~\ref{sec:performance} the experimental results are studied and interpreted.
Section~\ref{sec:conclusions} summarizes the paper and discusses future work.

\section{Background}\label{sec:background}
This section describes the main technologies leveraged in this research. In particular, 1) the malleability framework, DMR, 2) the reconfiguration engine, Proteo, 3) the resource manager, Slurm, and 4) the scientific application, MPDATA.

\subsection{DMRlib}
The Dynamic Management of Resources Library (DMRlib)~\cite{sergiothesis} is a high-level API that facilitates the adoption of malleability in HPC codes. DMRlib implements a communication layer between the parallel distributed runtime (PDR) and the RMS, making the management of processes and resources transparent to the user while providing great flexibility to increase productivity and resource utilization of HPC facilities~\cite{Iserte2020}.

Originally, DMRlib was implemented leveraging Nanos++ (the OmpSs runtime) and Slurm, as PDR and RMS respectively~\cite{Iserte2017}. DMRlib has been refactored in this work to support different PDR and RMS. Particularly, this paper presents DMR as a universal programming layer for dynamic resources and DMRlib as a modular communication layer among the application, the PDR, and the RMS.


DMRlib is communicated to Slurm via a customized resource selection plugin for dynamic resources, enabling resource reallocations in runtime.

DMR has proven to be a competitive malleability tool thanks to its various published success stories~\cite{Iserte2018hpg,Iserte2019a,reina_leon_implementacion_2024} and those currently in progress. 

\subsection{Proteo}\label{sec:background-proteo}

\textit{Proteo} is a highly configurable framework that allows the design of benchmarks to study the effect and the integration of malleability in real applications~\cite{proteo_2024}.
The framework is composed of two main modules: SAM (Synthetic Application Module), which is responsible for emulating the behavior of applications, and MaM (Malleability Module), which is responsible for incorporating malleability into real and emulated applications.
The combination of both allows the emulation of malleable parallel applications


MaM implements different variants to carry out two main tasks: process management and data redistribution, both defined as a combination of methods and strategies.

Regarding process management, we assume that the original group has $NP$ processes and the new group has $NT$ processes. The strategies provided are Baseline methods, which always spawn $NT$ processes,
and Merge methods, which recycles $min(NP, NT)$ processes in the original group, reducing the number of new processes spawned to $max(0, NT-NP)$.

Data must be moved from the original group to the new group, where variable data must be communicated synchronously, but constant data can also be communicated asynchronously. MaM implements point-to-point and collective methods as well as non-blocking and threaded communications.

\subsection{MPDATA Algorithm}

The MPDATA algorithm is the core of the multiscale fluid solver EULAG (Eulerian/semi-Lagrangian fluid solver) \cite{ROJ17a}. It is responsible for computing the advection of a nondiffusive quantity $\Psi$ within a flow field. Described by the continuity equation:

\begin{equation}
{{\partial \Psi} \over {\partial t}} + \text{div} ( \mathbf{V} \Psi ) = 0 ,
\end{equation}

where $\mathbf{V}$ represents the velocity vector. MPDATA employs a spatial discretization grounded in finite difference approximations~\cite{ROJ17b}.




Belonging to the forward-in-time class of algorithms, MPDATA executes iterative time steps, the count of which varies depending on the simulated physical phenomena. Each MPDATA step necessitates 5 input arrays and yields a single output array crucial for subsequent time steps.

Additionally to the GPU version of this malleable algorithm with DMR in~\cite{Iserte2019a}, in this work, we have implemented the malleable CPU version, which is more convenient for the testbed leveraged in the experiments \footnote{\url{https://gitlab.bsc.es/siserte/mpdata-dmr}}.

\section{Related Work}\label{sec:related}
In this section, we briefly review on--the--fly reconfiguration solutions integrated into a production-class resource managers to demonstrate the necessity of having a modular framework that supports different malleability engines from a system-wide perspective.

We enumerate the most relevant MPI solutions to enable dynamic processes and malleability. For more details, \cite{aliaga_survey_2022} and~\cite{ahmad} present a comprehensive survey on spawn-based process malleability and a more generalist dynamic resources state--of--the--art, respectively.

ReSHAPE~\cite{Sudarsan2007} is an integrated solution that adapts to varying workloads, comprising specialized reconfiguration libraries, a scheduler, and a runtime system. The tight integration of these components requires ReSHAPE users to create applications that align with this system.

The power-aware resource manager (PARM)~\cite{Sarood2014} employs over-provisioning, power capping, and job malleability to optimize job throughput within a stringent power budget in over-provisioned facilities. Malleability depends on the CHARM++ runtime support, which dynamically reallocates compute objects to processors.

The research presented in~\cite{Prabhakaran2015} integrates Adaptive Message Passing Interface (AMPI)~\cite{AMPI06}, which is an implementation of the MPI standard on top of Charm++, with the Torque/Maui job scheduler to address malleable jobs. This approach establishes a communication interface between the CHARM++ runtime and Torque/Maui.

Elastic MPI~\cite{compres_infrastructure_2016} comprises an infrastructure and a collection of API extensions designed for the malleable execution of MPI applications, leveraging Slurm and MPICH. By utilizing the functions offered by this API, an application can be designated as malleable, prompting the application's processes to verify if Slurm has initiated a reconfiguration periodically. Moreover, MPICH has been augmented with a new set of functions that supplement and replace the standard implementation. This approach is specific to MPICH, does not provide support for other MPI implementations, and does not facilitate data redistribution.


Besides spawn-based solutions, a new trend of dynamic process management is arising by leveraging modern MPI PSets, which are building blocks of the MPI Sessions concept introduced in version 4 of the MPI Standard. Although there is no consensus on implementing it among the different MPI distributions, some efforts have been made towards its expansion in the dynamic resources field, i.e., Parastation MPI and DPP~\cite{huber_towards_2022}.

Currently, there is a lack of a dynamic resource manager that spans both the parallel runtime and the resource manager. This manager should be compatible with standard MPI implementations and popular RMS systems. The Dynamic Resource Manager (DMR) is leading the way in standardizing dynamic resource management and making it more accessible by offering standard MPI support and integration with Slurm.

\section{DMR as a Common Friendly Programming Layer}\label{sec:dmr}
Originally, DMRlib was released as a framework to manage dynamic resources within an application, providing a friendly syntax. However, the recently revived interest in dynamic resource management has motivated several institutions to create their own framework. Since there is no standard and MPI does not plan to incorporate one in future releases, DMRlib has been transformed to support different dynamic solutions while source codes and users are agnostic to how resources are managed under the hood.

\begin{figure}[htb]
    \centering
    \includegraphics[width=0.95\linewidth]{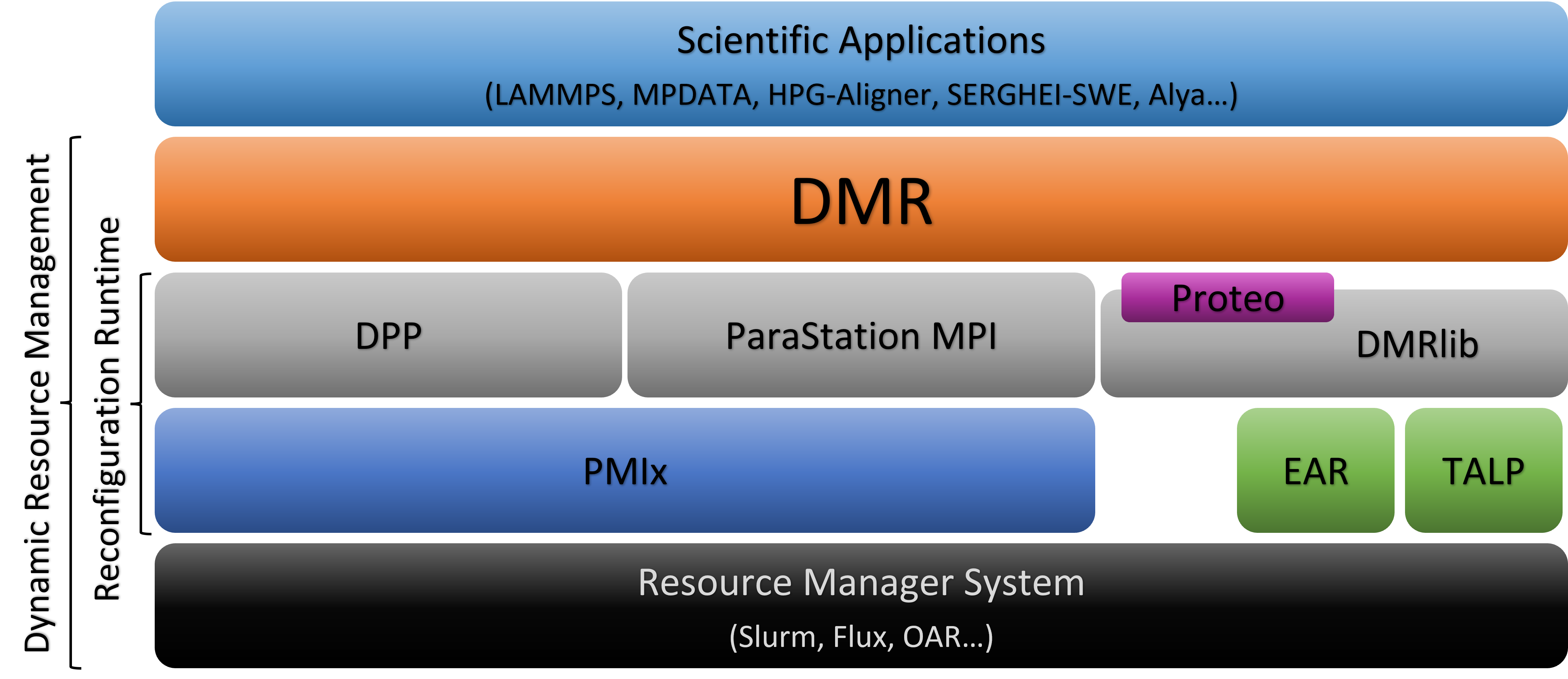}
    \caption{Dynamic Management of Resources Software Stack}
    \label{fig:dmr-layers}
\end{figure}

In particular, an excision of DMRlib has been created to provide a user API independent of the underlying dynamic resource manager. This excision is illustrated in Figure~\ref{fig:dmr-layers} in the layer ``DMR'', which is responsible for hiding the dynamic resources frameworks from the users.
In this regard, scientific applications can enable dynamicity by leveraging the new DMR API. Listing~\ref{code:dmr-macros} contains the mandatory wrappers to include in the user code.
These macros expect a series of user functions to perform the operations during the execution initialization and finalization (first and third, respectively) and the data redistribution functions, for instance, to send and receive in the cases of expansion and shrinkage.

\begin{lstlisting}[float, caption=DMR Macros., label=code:dmr-macros, captionpos=b, numbers=none]
DMR_INITIALIZE(initialize(...), redistribution(...));
DMR_RECONFIGURE(redistribution(...));
DMR_FINALIZE(finalize(...)); 
\end{lstlisting}

While \texttt{DMR\_INITIALIZE} and \texttt{DMR\_FINALIZE} can be invoked at the beginning and the end of the code, respectively, 
\texttt{DMR\_RECONFIGURATION}
is called whenever the execution is ready to be reconfigured, in other words, in synchronization points. For instance, iterative applications, such as physics simulations, present a main loop in which each iteration corresponds to a complete time step where the processes are synchronized previously to continue with the next step. That scenario is seen by dynamic resource management as a periodic reconfiguration point to trigger malleability.

If the dynamic resource manager determines to reconfigure the job, data must be redistributed accordingly to the user function ``redistribution''. For this purpose, this function is called when target processes have to initialize their data and update it to the current status of the execution. These data come from the parent processes that are sent during the reconfiguration stage.

In this work, we have extended DMR with an interoperability interface for the new version of DMRlib (and it has been done for Dynamic Processes with PSets (DPP)~\cite{posterISC24} and other ongoing work for other dynamic resources frameworks). Besides, DMRlib has been expanded with a pure-MPI backend able to operate without OmpSs, which was a strong dependency in the previous version~\cite{Iserte2020}. Although the logic is similar, DMRlib has substituted OmpSs pragma directives and Nanos++ runtime operations in favor of MPI standard functions, allowing the DMR interface for DMRlib to rely solely on MPI.

A brief example of this interface can be seen in Listing~\ref{code:dmr-dmrlib}. This listing showcases the interface for DMRlib of the finalization wrapper \texttt{DMR\_FINALIZE}, which executes as is the user function ``finalize''.

\begin{lstlisting}[float, caption=\textit{DMR\_FINALIZE} interface for DMRlib., label=code:dmr-dmrlib, captionpos=b, numbers=none]
#define DMR_FINALIZE(finalize) {                            \
    finalize;                                               \
}
\end{lstlisting}


The remaining interfaces are more intricate, and due to space constraints, they can be reviewed in the DMR repository\footnote{\url{https://gitlab.bsc.es/siserte/dmr}}.

MPDATA has been extended to support malleability leveraging the DMR API.
Listing~\ref{code:mpdata-dmr} showcases the main function of the malleable version of MPDATA.
It depicts a traditional iterative MPI application with the necessary logic to enable dynamic resources with DMR. 
Particularly, line~4 is responsible for initiating the environment after each reconfiguration.
Line~5 sets the limits of malleability by defining the lower and upper bound, the user-defined sweet spot of execution, and the stride for resource and reassignment,
respectively. 
The hardcoded values in the code are those used in the experimentation and are explained in Section~\ref{sec:performance}.
Furthermore, in line~6, the inhibitor will prevent two consecutive reconfiguration checks.
At the end of an iteration, reconfiguration can be triggered (line~9) with the corresponding data redistribution pattern when processes are synchronized.
Finally, memory is freed, and the program is terminated.

\begin{lstlisting}[float, caption=MPDATA main function., label=code:mpdata-dmr, captionpos=b]
void main(int argc, char **argv) {
  MPI_Init_thread(&argc, &argv, MPI_THREAD_MULTIPLE, NULL);
  init();                       // initialization
  DMR_INIT(mam_set(), mam_retrieve());
  DMR_Set_parameters(1, 4, 2, 2); 
  DMR_Inhibit_iter(2);
  while (DMR_it < TIME_STEPS) { // main loop
    update(); mpdata3d();       // halo exchange & computation
    DMR_RECONFIGURATION(mam_retrieve());
  }
  DMR_FINALIZE(release());      // finalize
  MPI_Finalize();
}
\end{lstlisting}

\section{Proteo Reconfiguration Engine in DMRlib}\label{sec:proteo}
DMRlib has been enhanced with Proteo's robust capabilities in process spawning and data redistribution mechanisms. Proteo now replaces the original reconfiguration engine in DMRlib, as illustrated in Figure~\ref{fig:dmr-layers}.

To achieve this goal, Proteo's interface has been incorporated into DMR, enabling Proteo to utilize DMRlib functionalities for communication with Slurm. 
This combination encompasses the adoption of performance-aware (via TALP~\cite{talp}) and energy-aware (via EAR~\cite{ear}) reconfiguration policies.

For example and analogously to the DMRlib interface presented in Listing~\ref{code:dmr-dmrlib}, Listing~\ref{code:dmr-proteo-fin} contains the \texttt{DMR\_FINALIZE} interface for Proteo.
In this interface, two tasks are performed: 1) to ensure that it is terminated if MaM does a non-blocking reconfiguration, and 2) to free the memory allocated by Proteo, DMR, and that specified by the user.
The rest of the interfaces can be found in the DMR repository.

Lines~4,~9, and~11 of Listing~\ref{code:mpdata-dmr} show how the DMR macros in Listing~\ref{code:dmr-macros} have been utilized to leverage Proteo spawning and data redistribution mechanisms.

Another upgrade provided by Proteo is in how the redistribution is handled. 
Proteo can redistribute 1D data with just a few hints from the user, easing the development task.
Listing~\ref{code:dmr-proteo-redist} shows the functions in MPDATA to indicate which data Proteo will redistribute.
Furthermore, \texttt{mam\_set} is invoked once by the first group of processes to declare the data pointers within Proteo runtime. 
Finally, \texttt{mam\_retrieve} is called by the resulting processes in a reconfiguration to get the new pointers for the data to use. 
It is important to note that these functions do not perform any communications, only declare/retrieve data for/from a redistribution.

\begin{lstlisting}[float, caption=\textit{DMR\_FINALIZE} interface for Proteo., label=code:dmr-proteo-fin, captionpos=b]
#define DMR_FINALIZE(finalize) {                            \
    if (DMR_state == MAM_PENDING)                           \
        MAM_Checkpoint(&DMR_state, MAM_WAIT_COMPLETION);    \
    finalize;                                               \
    MAM_Finalize();                                         \
    if (DMR_COMM != MPI_COMM_WORLD && DMR_COMM != NULL)     \
        MPI_Comm_disconnect(&DMR_COMM);                     \
}
\end{lstlisting}

\begin{lstlisting}[float, caption=Redistribution functions for Proteo., label=code:dmr-proteo-redist, captionpos=b]
void mam_set() {
  real *redist_arrays[] = {x_, u1_, u2_, u3_};
  int total_size = n * (m + 1) * (l + 1) * DMR_comm_size;
  int haloSize = HALO * (m + 1) * (l + 1);
  for (int i=0; i<4; i++) {
    MAM_Data_add(redist_arrays[i] + haloSize, NULL, total_size, mpi_real, MAM_DATA_DISTRIBUTED, MAM_DATA_VARIABLE);
  }
}
void mam_retrieve() {
  release(); init(); // Free previous and allocate new memory
  real *aux_array, *redist_arrays[] = {x_, u1_, u2_, u3_};
  int localSize = n * (m + 1) * (l + 1);
  int haloSize = HALO * (m + 1) * (l + 1);
  for (int i=0; i<4; i++) {
    MAM_Data_get_pointer(&(aux_array), i, NULL, NULL, MAM_DATA_DISTRIBUTED, MAM_DATA_VARIABLE);
    memcpy(redist_arrays[i] + haloSize, aux_array, size_real*localSize)
  }
}
\end{lstlisting}

\section{Experimental Results}\label{sec:performance}
The experiments were executed on a cluster of eight servers, each equipped with two 10-core Intel Xeon 4210 processors. The nodes are interconnected via EDR Infiniband at 100Gb/s. 
The MPI version used is MPICH 4.2.1~\footnote{\url{https://www.mpich.org}}, compiled with CH4:OFI-embedded netmod (Infiniband).
A custom version of Slurm\footnote{\url{https://gitlab.bsc.es/siserte/dmr/-/tree/mam-mn5}} was leveraged in this study to support process malleability and dynamic reconfiguration of jobs.

\subsection{Reconfiguration Scheduling Policy}
In addition to traditional job submission, where jobs ask for a specific number of resources, we have enabled Slurm's flexible submission option, which allows a job to request a range of resources (aka moldability).
The proposed reconfiguration policy is designed to optimize overall productivity by maximizing the number of completed jobs per unit of time and is executed by Slurm when any job triggers a reconfiguration.
This approach involves requesting jobs to provide information on their lower and upper reconfiguration limits and a ``sweet spot'' defined by the user. Algorithm~\ref{alg:slurm} outlines the implementation of this policy. 
The reconfiguration stride adheres to power of two values as $[\forall x \in X, \exists k \in \mathbb{N}, x = 2^k
]$.

When a job initiates a reconfiguration, Slurm checks the queue (line~1).
The job is expanded if it runs alone in the cluster and has available resources (line~2).
If there are pending jobs, the invoker job runs with fewer resources than its preferred configuration, and when there are available resources, the job is expanded (lines~3-5). 

A job can only be downsized if its current resource allocation exceeds the preferred value, and a pending job could be initiated thanks to the freed resources (lines~6-7). 
Subsequently, the Slurm's priority of the target job to be started is increased.

The job is expanded when no pending job can be started, and spare resources are available (lines 8-9).

\begin{algorithm}[htb]
  \scriptsize
\caption{Reconfiguration policy algorithm in Slurm}\label{alg:slurm}
\begin{algorithmic}[1]
\If{\Not{$pending\_jobs()$}}
    \If {$avail\_resources()$}
        \Return {expand}
    \EndIf
\Else 
    \If {$current < preferred$}
        \If {$avail\_resources()$}
            \Return {expand}
        \EndIf
    \Else
        \If{$pending\_job\_can\_be\_initiated()$}
            increase\_pending\_job\_priority()
   	    \Return {shrink}
        \Else 
            \If {$avail\_resources()$}
                \Return {expand}
            \EndIf
        \EndIf
    \EndIf
\EndIf
\Return{none}
\end{algorithmic}
\end{algorithm}

\subsection{Workload Configuration}
MPDATA instances are submitted to Slurm in the shape of jobs.
All the jobs are identically configured with a domain of size $1024x128x32$ and 20 time steps.
Malleability is configured as line~5 of Listing~\ref{code:mpdata-dmr} shows, defining $1$ and $4$ nodes as the lower and upper bound, $2$ nodes as the sweet spot, and limiting reconfigurations to doubling and halving. 
These values were calculated after a scalability analysis of MPDATA, that was omitted in this paper because of length restrictions.
Furthermore, jobs are submitted moldable, letting Slurm initiate jobs with $\{1,2,4\}$ nodes.
Jobs are executed with 16 MPI ranks per node and a single OpenMP thread per rank.

The workload comprises $1,000$ MPDATA jobs, which are submitted with an inter-arrival time of one second to simulate a stressed scenario, one of the most unfavorable situations in dynamic resources management.

Proteo has been configured to perform the redistribution with synchronous point-to-point communications, while the process management is handled with either the synchronous Baseline method or the synchronous Merge method with the strategy \textit{multiple}. The \textit{multiple} strategy ensures that Slurm's restrictions are satisfied for the Merge method.

\subsection{Evaluation}
\begin{figure}[htb]
    \centering
    \includegraphics[clip,width=0.75\textwidth,trim={0.1cm 1cm 0.1cm 0.1cm}]{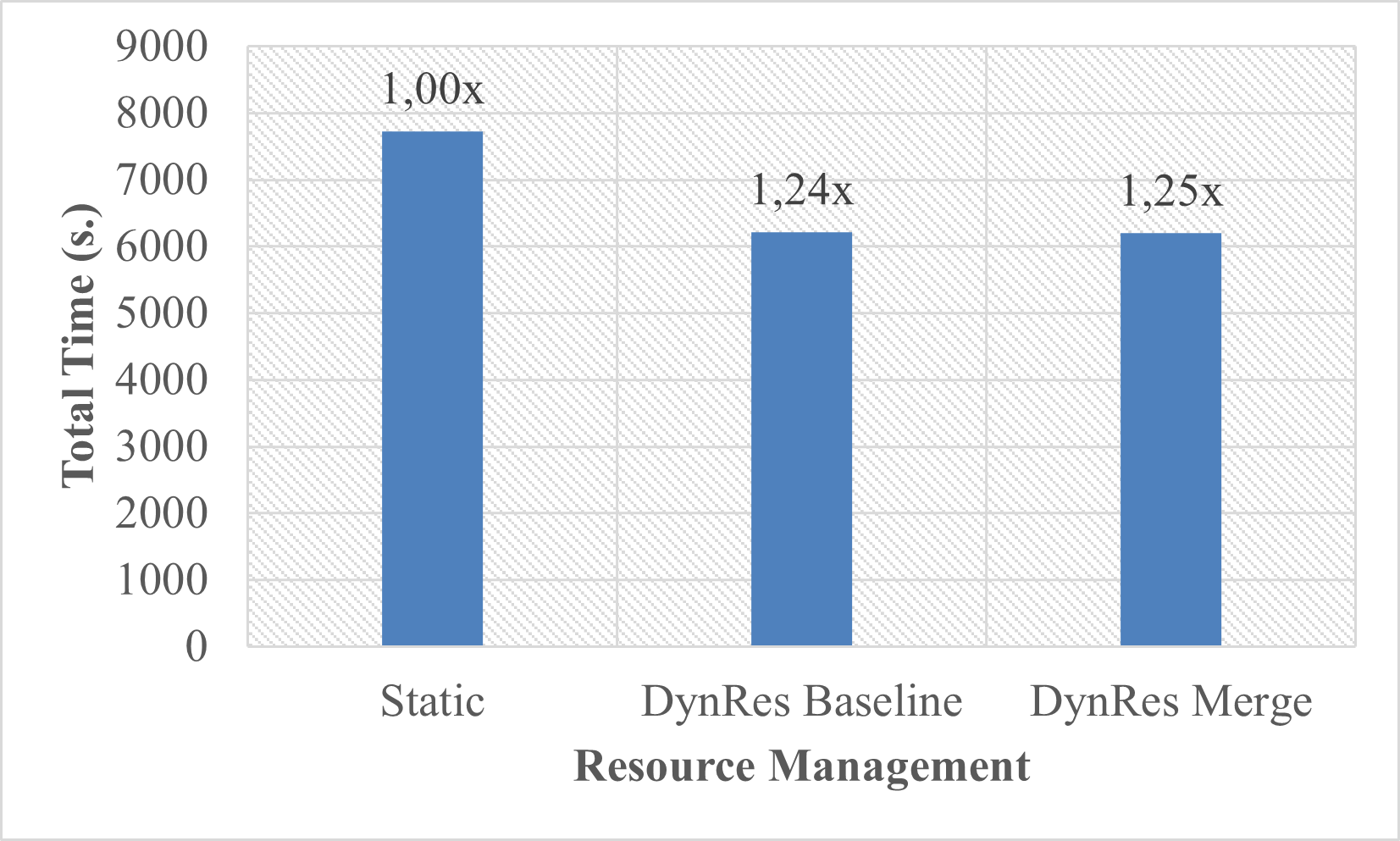}
    \caption{Workload Completion Times.}
    \label{fig:workload-times}
\end{figure}

Results show a reduced workload completion time when comparing the dynamic workloads to the static one.
Figure~\ref{fig:workload-times} depicts the times, in seconds, along the speedups (bar labels) compared to the traditional static resource management.
In this regard, dynamic resources (``DynRes'') workloads are distinguished between respawning all the processes (Baseline) and spawning just the additional processes or removing those not needed anymore (Merge) techniques explained in Section~\ref{sec:background-proteo}.
Notice that the \textit{Baseline} strategy was also implemented in DMRlib but without support for multiple processes per node. In this regard, DMRlib and Proteo cannot be directly compared because they are not competitors but complementaries.

Comparing the execution time between the baseline and merge versions, there is no significant difference in performance. There are two related reasons for this conclusion: First, each job performs at most one reconfiguration in its lifetime, and the difference between the methods becomes more noticeable when there are more reconfigurations~\cite{proteo_2024}. Second, the way in which malleability is implemented forces the use of the baseline version in the first reconfiguration to ensure that Slurm's restrictions regarding release nodes are satisfied.

Table~\ref{tab:exec-times} contains the mean time of all the jobs in the workload for their execution, waiting, and completion (execution + waiting) times, respectively.
The slowdown higher than 1.5x in the job execution times is compensated by the speedup of the same factor when it comes to mean waiting time, which in turn is the dominant time in the completion time of a job.

\begin{table}[htb]
    \centering
    \begin{tabular}{r|c|c|c}
         & \text{Execution Mean Time} & \text{Waiting Mean Time} & \text{Completion Mean Time} \\ \hline 
         \text{Static} & 14.67 \text{ s.} & 3,356.32 \text{ s.} & 3,370.99 \text{ s.} \\ 
        \text{DynRes Baseline} & 23.52 \text{ s.} & 2,516.51 \text{ s.} & 2,540.03 \text{ s.} \\ 
        \text{DynRes Merge} & 23.51 \text{ s.} & 2,541.13 \text{ s.} & 2,564.64 \text{ s.} \\ \hline
    \end{tabular}
    \caption{Mean job times for different resource management techniques.}
    \label{tab:exec-times}
\end{table}

Table~\ref{tab:allocation} represents the percentage of allocated time during the workload execution per node ([n01-n08]). These data are captured by Slurm, and overall, dynamic experiments show higher utilization rates than the static counterpart.
4-node jobs in the static configuration perfectly match an 8-node cluster (two jobs running concurrently). Although this scenario is one of the most beneficial for the static setup, dynamic resources can even beat it thanks to adapting jobs at the beginning and end of the workload when there is more resource fragmentation.

\begin{table}[htb]
    \centering
    \begin{tabular}{r|c|c|c|c|c|c|c|c}
        & n01 & n02 & n03 & n04 & n05 & n06 & n07 & n08 \\ \hline
        Static & 89.19\% & 89.19\% & 89.19\% & 89.19\% & 87.64\% & 87.64\% & 87.64\% & 87.64\% \\ \hline
        DynRes Baseline & 93.18\% & 93.10\% & 93.02\% & 93.94\% & 91.88\% & 91.83\% & 91.38\% & 91.56\% \\ \hline
        DynRes Merge & 93.01\% & 93.33\% & 94.01\% & 93.56\% & 91.38\% & 91.31\% & 91.80\% & 91.81\% \\ \hline
    \end{tabular}
        \caption{Utilization time per node and experiment.}
    \label{tab:allocation}
\end{table}

\section{Conclusions}\label{sec:conclusions}
This paper is born from the need for a dynamic resource manager that relies on the standard MPI, which supports different reconfiguration strategies, is compatible with a ubiquitous HPC RMS manager such as Slurm, and is easy to use when developing scientific applications. 

In the state--of--the--art, we cannot find a dynamic resource manager transversal to the parallel runtime and the resource manager, which relies on standard MPI implementations and widely used RMS. DMR is paving the road towards standardizing dynamic resource management and democratization, providing standard MPI support and interacting with Slurm.

In this regard, Proteo interfaces for DMR have been implemented; Leveraging DMR, an MPDATA malleable version, has been developed; Proteo and DMRlib have established a synergy to support Slurm and additional spawning strategies, respectively.

Results show that with little programming effort, a workload can benefit from the dynamic management of resources in a cluster. The same workload can be processed with 75\% of the original time and a high utilization rate when enabling dynamic resources even in the most beneficial scenario for staticisty.

Ongoing research with extended experimental setups will provide more comprehensive insights, including asynchronous reconfiguration in Proteo, more variety of job types and different inter-arrival, larger computing infrastructure such as Marenostrum 5, and a comprehensive study of metrics such as resource utilization, energy consumption, and job/workload times.

\section*{Acknowlegments}
The authors would like to thank the ``BSC Marenostrum Hackathon'' series for fostering this collaboration.

The researchers from BSC are involved in the project The European PILOT, which has received funding from the European High-Performance Computing Joint Undertaking (JU) under grant agreements No. 101034126 and No. PCI2021-122090-2A under the MCIN/AEI and the EU NextGenerationEU/PRTR.
They are also grateful for the support from the Department of Research and Universities of the Government of Catalonia to the AccMem (Code: 2021 SGR 00807).
Antonio J. Peña was partially supported by the Ramón y Cajal fellowship RYC2020-030054-I funded by MCIN/AEI/ 10.13039/501100011033 and by ``ESF Investing in your future''.

The researchers from UJI have been funded by project PID2020-113656RB-C21 supported by MCIN/AEI/10.13039/501100011033.
Researcher I.~Martín-Álvarez was supported by the predoctoral fellowship ACIF/2021/260 from Valencian Region Government and European Social Funds.

%
%
\bibliographystyle{splncs04} 
\bibliography{CP039}

@article{ahmad,
  author    = {Tarraf, Ahmad and et al.},
  title     = {{Malleability in Modern HPC Systems: Current
Experiences, Challenges, and Future Opportunities}},
  journal   = {IEEE Transaction on Parallel and Distributed Systems},
  year      = {2024},
  note       = {(in-press)},
}

@article{ROJ17a,
  author    = {Krzysztof Rojek and Roman Wyrzykowski},
  title     = {{Performance modeling of 3D MPDATA simulations on GPU cluster}},
  journal   = {The Journal of Supercomputing},
  year      = {2017},
  volume    = {73},
  number    = {2},
  pages     = {664--675},
  doi       = {10.1007/s11227-016-1774-z},
}

@article{ROJ17b,
author = {Rojek, Krzysztof and Wyrzykowski, Roman and Kuczynski, Lukasz},
title = {{Systematic adaptation of stencil-based 3D MPDATA to GPU architectures}},
journal = {Concurrency and Computation: Practice and Experience},
volume = {29},
number = {9},
pages = {e3970},
doi = {10.1002/cpe.3970},
year = {2017}
}

@inproceedings{posterISC24,
title={{Leveraging Dynamic Resource Management in HPC}},
author={P. Dutot and J. Fecht and K. Gaddameedi and D. Huber and S. Iserte and M. Minion and M. Schulz and M. Schreiber and V. Schüller and A. J. Peña and O. Richard},
booktitle = {ISC High Performance},
	month = jun,
	year = {2024},
	isbn = {pending}
}

@article{ear,
  title={EAR: Energy Management Framework for Supercomputers},
  author={Corbalan, Julita and Brochard, Luigi},
  journal={Barcelona Supercomputing Center (BSC)},
  year={2019}
}

@inproceedings{talp,
author = {Lopez, Victor and Ramirez Miranda, Guillem and Garcia-Gasulla, Marta},
title = {{TALP: A Lightweight Tool to Unveil Parallel Efficiency of Large-Scale Executions}},
year = {2021},
isbn = {9781450383875},
doi = {10.1145/3452412.3462753},
booktitle = {Proceedings of PERMAVOST},
}

@article{reina_leon_implementacion_2024,
	title = {Implementación distribuida maleable del método Laplace},
	url = {https://openaccess.uoc.edu/handle/10609/149763},
	author = {Reina León, José},
        year = {2024},
	note = {{UOC}},
}

@inproceedings{huber_towards_2022,
	title = {{Towards Dynamic Resource Management with {MPI} Sessions and {PMIx}}},
	isbn = {978-1-4503-9799-5},
	doi = {10.1145/3555819.3555856},
	pages = {57--67},
	booktitle = {Proceedings of the 29th EuroMPI/USA},
	author = {Huber, Dominik and Streubel, Maximilian and Comprés, Isaías and Schulz, Martin and Schreiber, Martin and Pritchard, Howard},
	urldate = {2024-03-04},
	date = {2022-09-14},
        year = {2022},
}

@article{aliaga_survey_2022,
	title = {A {Survey} on {Malleability} {Solutions} for {High}-{Performance} {Distributed} {Computing}},
	volume = {12},
	copyright = {http://creativecommons.org/licenses/by/3.0/},
	issn = {2076-3417},
	doi = {10.3390/app12105231},
	language = {en},
	number = {10},
	urldate = {2024-01-23},
	journal = {Applied Sciences},
	author = {Aliaga, Jose I. and Castillo, Maribel and Iserte, Sergio and Martín-Álvarez, Iker and Mayo, Rafael},
	month = jan,
	year = {2022},
	pages = {5231},
}

@article{proteo_2024,
	title = {Proteo: A Framework for the Generation and Evaluation of Malleable {MPI} Applications},
	journal = {The Journal of Supercomputing},
	author = {Martín-Álvarez, Iker and Aliaga, Jose I. and Castillo, Maribel and Iserte, Sergio},
	year = {in 2nd revision}
}

@inproceedings{compres_infrastructure_2016,
	series = {{EuroMPI} 2016},
	title = {Infrastructure and {API} {Extensions} for {Elastic} {Execution} of {MPI} {Applications}},
	isbn = {978-1-4503-4234-6},
	doi = {10.1145/2966884.2966917},
	booktitle = {Proceedings of the 23rd {EuroMPI}},
	author = {Comprés, Isaías and Mo-Hellenbrand, Ao and Gerndt, Michael and Bungartz, Hans-Joachim},
	year = {2016},
	pages = {82--97},
}

@INPROCEEDINGS{AMPI06,
        author="Chao Huang and Gengbin Zheng and Sameer Kumar and Laxmikant V. Kal\'{e}",
        title="{Performance Evaluation of Adaptive MPI}",
	booktitle = "Proceedings of ACM SIGPLAN Symposium on Principles and Practice of Parallel Programming 2006",
	month = "March",
	year = "2006"
}

@article{Iserte2019a,
author = {Iserte, Sergio and Rojek, Krzysztof},
doi = {10.1007/s11227-019-03034-x},
issn = {0920-8542},
journal = {The Journal of Supercomputing},
month = {oct},
pages = {1--20},
publisher = {Springer US},
title = {{An study of the effect of process malleability in the energy efficiency on GPU-based clusters}},
year = {2019}
}

@article{Iserte2018hpg,
author = {Iserte, Sergio and Mart{\'{i}}nez, H{\'{e}}ctor and Barrachina, Sergio and Castillo, Maribel and Mayo, Rafael and Pe{\~{n}}a, Antonio J},
doi = {10.1177/1094342018802347},
issn = {1094-3420},
journal = {The International Journal of High Performance Computing Applications},
month = {sep},
publisher = {SAGE PublicationsSage UK: London, England},
title = {{Dynamic reconfiguration of noniterative scientific applications}},
year = {2018}
}

@article{Iserte2020,
author = {Iserte, S. and Mayo, R. and Quintana-Orti, E.S. and Pena, A.J.},
doi = {10.1109/TC.2020.3022933},
issn = {15579956},
journal = {IEEE Transactions on Computers},
title = {{DMRlib: Easy-coding and Efficient Resource Management for Job Malleability}},
year = {2020}
}

@inproceedings{Iserte2017,
address = {Bristol (UK)},
author = {Iserte, Sergio and Mayo, Rafael and Quintana-Ort{\'{i}}, Enrique S. and Beltran, Vicen{\c{c}} and Pe{\~{n}}a, Antonio J.},
booktitle = {46th International Conference on Parallel Processing Workshops},
doi = {10.1109/ICPPW.2017.36},
isbn = {978-1-5386-1044-2},
month = {aug},
pages = {180--189},
publisher = {IEEE},
title = {{Efficient Scalable Computing through Flexible Applications and Adaptive Workloads}},
year = {2017}
}

@phdthesis{sergiothesis,
address = {Castell{\'{o}} de la Plana},
author = {Iserte, Sergio},
doi = {10.6035/14101.2018.176272},
month = {nov},
school = {Universitat Jaume I},
title = {{High-throughput Computation through Efficient Resource Management}},
year = {2018}
}

@inproceedings{Prabhakaran2015,
author = {Prabhakaran, Suraj and Neumann, Marcel and Rinke, Sebastian and Wolf, Felix and Gupta, Abhishek and Kale, Laxmikant V.},
booktitle = {IEEE IPDPS},
isbn = {978-1-4799-8649-1},
month = may,
pages = {429--438},
title = {{A Batch System with Efficient Adaptive Scheduling for Malleable and Evolving Applications}},
year = {2015}
}

@inproceedings{Sarood2014,
author = {Sarood, Osman and Langer, Akhil and Gupta, Abhishek and Kale, Laxmikant},
booktitle = {SC14: International Conference for High Performance Computing, Networking, Storage and Analysis},
isbn = {978-1-4799-5500-8},
month = {nov},
pages = {807--818},
publisher = {IEEE},
title = {{Maximizing Throughput of Overprovisioned HPC Data Centers Under a Strict Power Budget}},
year = {2014}
}

@inproceedings{Sudarsan2007,
author = {Sudarsan, Rajesh and Ribbens, Calvin J.},
booktitle = {Proceedings of the International Conference on Parallel Processing},
isbn = {076952933X},
title = {{ReSHAPE: a Framework for Dynamic Resizing and Scheduling of Homogeneous Applications in a Parallel Environment}},
year = {2007}
}
%
\end{document}